\documentclass[
   final            % use final for the camera ready runs
  ]
  {aipproc}

\layoutstyle{6x9}
\usepackage{graphicx}% Include figure files
\usepackage{dcolumn}%Align table columns on decimal point
\usepackage{amsmath}
\usepackage{amsfonts}
\usepackage{amssymb}
\usepackage{bm}% bold math
\begin{document}

\title{Sensitivity of the electric dipole polarizability to the neutron skin thickness in ${}^{208}$Pb}



\author{X.~Roca-Maza}
{address={INFN, sezione di Milano, via Celoria 16, I-20133 Milano, Italy}}

\author{B. K. Agrawal}
{address={Saha Institute of Nuclear Physics, Kolkata 700064, India}}

\author{G. Col\`o}
{address={Dipartimento di Fisica, Universit\`a degli Studi di Milano},
altaddress={INFN,  Sezione di Milano, 20133 Milano, Italy}}

\author{W. Nazarewicz}
{address={Department of Physics and Astronomy, University of Tennessee, Knoxville, Tennessee 37996, USA},
altaddress={Physics Division, Oak Ridge National Laboratory, Oak Ridge, Tennessee 37831, USA},
altaddress={Institute of Theoretical Physics, University of Warsaw, Hoza 69, PL-00-681 Warsaw, Poland}}

\author{N. Paar}
{address={Physics Department, Faculty of Science, University of Zagreb, Zagreb, Croatia}}

\author{J. Piekarewicz}
{address={Department of Physics, Florida State University, Tallahassee, FL 32306, USA}}

\author{P.-G. Reinhard}
{address={Institut f\"ur Theoretische Physik II, Universit\"at Erlangen-N\"urnberg, Staudtstrasse 7, D-91058 Erlangen, Germany}}

\author{D. Vretenar}
{address={Physics Department, Faculty of Science, University of Zagreb, Zagreb, Croatia}}

\begin{abstract}
The static dipole polarizability, $\alpha_{\rm D}$, in ${}^{208}$Pb has been recently measured with high-resolution via proton inelastic scattering at the Research Center for Nuclear Physics (RCNP) \cite{tami11}. This observable is thought to be intimately connected with the neutron skin thickness, $r_{\rm skin}$, of the same nucleus and, more fundamentally, it is believed to be associated with the density dependence of the nuclear symmetry energy. The impact of $r_{\rm skin}$ on $\alpha_{\rm D}$ in ${}^{208}$Pb is investigated and discussed on the basis of a large and representative set of relativistic and non-relativistic nuclear energy density functionals (EDF) \cite{piek12}. 
\end{abstract}

\maketitle
%%%%%%%%%%%%%%%%%%%%%%%%%%%%%%%%%%%%%%%%%%%%
%% MAINMATTER
%%%%%%%%%%%%%%%%%%%%%%%%%%%%%%%%%%%%%%%%%%%%
\section{Introduction}

The Lead Radius Experiment (PREX) \cite{hor01a,mich05} has recently measured $r_{\rm skin}$, defined as the difference between the neutron and proton root mean square radii, of ${}^{208}$Pb \cite{abra12}. This experiment is performed via parity-violating electron scattering \cite{donn89} and provides the first purely electroweak measurement of the neutron distribution of a heavy nucleus. The neutron skin is strongly dependent on the isovector properties of nuclei and impacts on a variety of areas such as nuclear structure \cite{brow00,furn02,cent09,rein10,roc11a}, atomic parity violation \cite{poll92}, and neutron-star structure \cite{hor01b,stei05}. By measuring the neutron form factor of ${}^{208}$Pb at $q\approx 0.475$ fm${}^{-1}$, PREX was able to determine $r_{\rm skin} = 0.33^{+0.16}_{-0.18}$ fm \cite{abra12}.

Alternatively, although the estimation of the neutron distribution in nuclei based on measurements using hadronic probes are model dependent and display large theoretical uncertainties \cite{ray85,ray92}, the use of these probes for the direct or indirect determination of such an observable is nowadays growing due to the necessity of improving our knowledge in the isovector channel of the nuclear effective interaction \cite{trcz01,klos07,brow07,clar03,zeni10}. The analysis from recent experiments have led to values for $r_{\rm skin} = 0.16 \pm (0.02)_{\rm stat} \pm (0.04)_{\rm syst}$ fm \cite{klos07} and $r_{\rm skin} = 0.211^{+0.054}_{-0.063}$ fm \cite{zeni10} in ${}^{208}$Pb. 

The electric dipole polarizability, $\alpha_{\rm D}$, is another observable sensitive to the isovector properties of the nuclear effective interaction. This quantity is obtainable from the linear response of the system to an external dipolar field of the form, $F_{\rm D} = (Z/A)\sum_i^N r_n Y_{1M}(\hat{r}_n) - (N/A)\sum_i^Z r_p Y_{1M}(\hat{r}_p)$, being $N$, $Z$, $A$, $r_j$ with $j= n$ or $p$ and $Y(\hat{r}_j)$ the neutron, proton and mass numbers, the radial position of the $j-$th nucleon and the spherical harmonic, respectively. If $\vert 0\rangle$ is the ground state and $\vert\nu\rangle$ is a complete set of excited states, the polarizability can be written as follows, $\alpha_{\rm D} = (8\pi/9) e^2 m_{-1} = (8\pi/9) e^2 \sum_\nu (\vert\langle \nu\vert F_{\rm D}\vert 0\rangle\vert^2/\omega_\nu)$ where $m_{-1}$ is the inverse energy weighted sum rule or inverse energy moment of the strength function, $R_{E1}(\omega)=\sum_\nu \vert\langle \nu\vert F_{\rm D}\vert 0\rangle\vert^2\delta(\omega-\omega_\nu)$ which evaluates the dipole response. For stable medium and heavy nuclei, the dipole response is largely concentrated in the giant dipole resonance (GDR) \cite{hara01}. In this isovector mode, commonly viewed as an oscillation of neutrons against protons, the symmetry energy at some sub-saturation density acts as the restoring force \cite{trip08}. In addition, it is important to notice that the possible presence of a low-lying dipole strength may have a non negligible effect on $\alpha_{\rm D}$ \cite{piek11,roc11b,vret12}.     

Actually, $r_{\rm skin}$ is expected to be linearly correlated with $\alpha_{\rm D}$ based on both macroscopic arguments \cite{lipp82,satu06} and microscopic calculations \cite{rein10,piek11}. The high-resolution measurement at RCNP of the $E1$ strength distribution $R_{E1}(\omega)$, where $\omega$ is the excitation energy, in ${}^{208}$Pb \cite{tami11} has allowed to deduce the experimental value of $\alpha_{\rm D} = 20.1 \pm 0.6$ fm${}^{3}$. Actually, Tamii {\it et al.} \cite{tami11}, relying on the predictions of one single EDF \cite{klup09} deduced a value of $r_{\rm skin}=0.156^{+0.025}_{-0.021}$ fm for ${}^{208}$Pb. However, systematic errors were not estimated. Motivated by the interesting physics behind this observable, we present an exhaustive analysis of the correlation between $\alpha_{\rm D}$ and $r_{\rm skin}$ in ${}^{208}$Pb within a large set of EDFs (see Ref.~\cite{piek12} for more details).

\section{Results}

In this contribution, we present our recent results on the correlation between $r_{\rm skin}$ and $\alpha_{\rm D}$ in ${}^{208}$Pb using a representative set of both relativistic and non-relativistic EDFs \cite{piek12}. In all cases, these self-consistent models have been calibrated to selected global properties of finite nuclei and infinite nuclear matter. These models have been used without any further adjustment to compute $R_{E1}$ using the consistent random-phase approximation (RPA). 

\begin{figure}
  \includegraphics[width=0.5\textwidth]{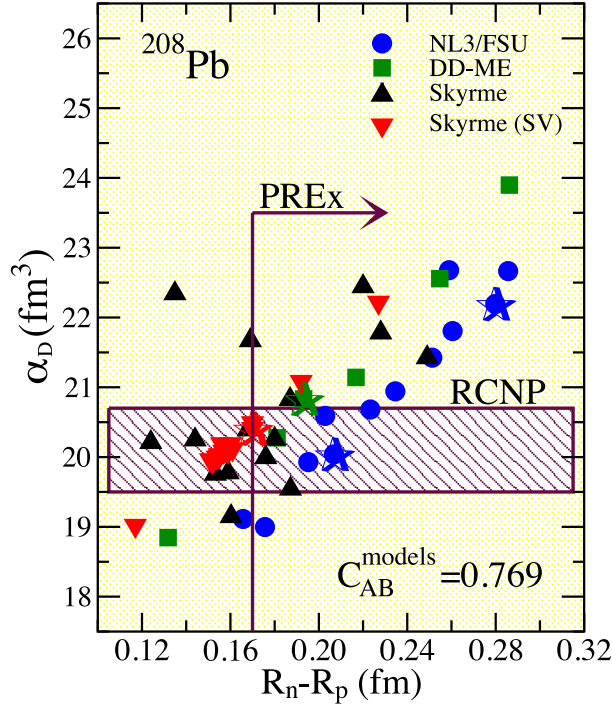}
  \caption{$\alpha_{\rm D}$ and $r_{\rm skin}$ of ${}^{208}$Pb as predicted by 48 nuclear EDFs (see \cite{piek12} for more details). Constrains on $r_{\rm skin}$ from PREX \cite{abra12} and on $\alpha_{\rm D}$ from RCNP \cite{tami11} are also shown.}
\label{fig1}
\end{figure}

We have chosen 48 EDFs and show their predictions in Fig.~\ref{fig1} (for details and original references see Ref.~\cite{piek12}). The up triangles correspond to Skyrme EDFs that have been widely used in the literature and fitted using very different protocols. In addition, we found interesting to analyze the predictions of three different families of relativistic (squares) and Skyrme EDFs (circles and down triangles) in which the value of the symmetry energy have been systematically varied around an optimal model (depicted by a star in Fig.~\ref{fig1}). All the models within a family remain still accurate although their departure from the optimal model. This is basically due to the fact that the isovector channel of the nuclear interaction is not tightly constrained by available data. 

Although some scatter is shown in Fig.~\ref{fig1}, the approximate linear relation between $\alpha_{\rm D}$ and $r_{\rm skin}$ discussed above is roughly confirmed by the different models. Specifically, the predicted linear correlation coefficient is around 0.8 when all the 48 models are taken into account while it is close to one when specific families of interactions are considered separately. The exception to this rule comes from the subset of Skyrme interactions that do not belong to a family of systematically varied interactions (black circles). This points to the fact that other quantities that do not appreciably vary in the different families of interactions are, indeed, affecting the value of $\alpha_{\rm D}$. It might be the case of some (isoscalar) properties that are almost constant within each one of the families and change when one looks at different families. This should be further investigated.  

In Fig.~\ref{fig1}, the constrains on $r_{\rm skin}$ from PREX \cite{abra12} and on $\alpha_{\rm D}$ from RCNP \cite{tami11} are also shown. By looking at the models that are compatible with the RCNP measurement, we perform an average of our theoretical results and obtain a $r_{\rm skin} = 0.168 \pm 0.022$ fm. Almost all theoretical predictions that agree with $\alpha_{\rm D}$ (within the experimental error bars) are consistent with the value of $r_{\rm skin}$ measured by PREX. 

\section{Summary and conclusions}

We have analyzed the correlation between $\alpha_{\rm D}$ and $r_{\rm skin}$ in ${}^{208}$Pb using a large set of representative EDFs. Macroscopic analyses suggest that these two observables should be correlated. We have seen that in our study, that within families of accurately calibrated models a strong correlation between $r_{\rm skin}$ and $\alpha_{\rm D}$ in ${}^{208}$Pb arise. When these models are combined and more differences appear between them, the correlation weakens. We have estimated the model or theoretical systematic error of $r_{\rm skin}$ compatible with the measurement at RCNP and found that the obtained value is compatible but still far from the central value obtained by PREX.

%%%%%%%%%%%%%%%%%%%%%%%%%%%%%%%%%%%%%%%%%%%%%%%%
%% BACKMATTER
%%%%%%%%%%%%%%%%%%%%%%%%%%%%%%%%%%%%%%%%%%%%%%%%

\begin{theacknowledgments}
Work partially supported by the Italian Research Project ``Many-Body Theory of Nuclear Systems and Implications on the Physics of Neutron Stars'' (PRIN 2008), by the Office of Nuclear Physics, U.S. Department of Energy under Contract Nos. DE-FG05-92ER40750 (FSU), DEFG02-96ER40963 (UTK); and by the BMBF under Contract 06ER9063. 
\end{theacknowledgments}

%%%%%%%%%%%%%%%%%%%%%%%%%%%%%%%%%%%%%%%%%%%%%%%%
%% BIBLIOGRAPHY
%%%%%%%%%%%%%%%%%%%%%%%%%%%%%%%%%%%%%%%%%%%%%%%%

\bibliographystyle{aipproc}   % if natbib is available

\end{document}